\documentclass{roadef2012}

\begin{document}
\vspace{-0.3cm}
\title{La théorie des jeux pour l'établissement des contrats dans les réseaux interdomaines}
\vspace{-0.3cm}
\def\shorttitle{Titre court}
\vspace{-0.3cm}
\author{Dominique Barth, Boubkeur Boudaoud et Thierry Mautor}

\institute{
Université de Versailles Saint Quentin \\ 45 Avenue des états unis, 78035, Versailles (France)\\
\email{ \{dominique.barth,boubkeur.boudaoud,thierry.mautor\}@prism.uvsq.fr}
}

\maketitle
\thispagestyle{empty}
\vspace{-0.35cm}
\section{Introduction}
\vspace{-0.2cm}
Dans ce travail, nous montrons comment mettre en \oe{}uvre la gestion des ressources et la garantie de QoS (Quality of Service) dans l'interdomaine en utilisant le \textbf{modèle en stock} \cite{BBT11}. Dans \cite{BBT11}, nous avons appliqué un modèle distribué pour l'établissement des SLAs (Service Level Agreement) entre les opérateurs pour l'achat des routes avec une garantie de QoS et des résultats significatifs ont été obtenus sur des topologies simples. Dans ce travail, nous appliquons un modèle de jeu pour l'établissement des SLAs. Dans le modèle en stock  \cite{BBT11}, nous avons considéré que les clients achètent chez leurs voisins des routes  avec des garanties de QoS. Cet achat se fait à travers la passation d'un contrat établi entre le client et son fournisseur pour la capacité en stock réservée sur une route à destination donnée. Cette \textbf{capacité temporelle} est disponible pendant une durée bien précise définie en unité de temps. Chaque domaine souhaitant acheter une route vers une destination donnée, doit négocier avec son voisin (fournisseur) pour réserver une certaine capacité dans l'intervalle $[cap\_min, cap\_max]$ sur cette route. La raison principale d'introduire cet intervalle de capacité est de permettre au client l'ajustement de sa capacité. Chaque route avec sa source et sa destination, correspond à un ensemble de contrats entre chaque paire d'opérateurs voisins, de la source à la destination. Un n\oe{}ud peut proposer une route à ses voisins seulement s'il a une capacité disponible en stock sur cette route. C'est dans ce cadre qu'on parle de modèle en \textbf{cascade inversée} présenté dans \cite{BBT11}.
\vspace{-0.35cm}
\section{Modèle de jeu pour l'achat des routes } 
\vspace{-0.2cm}
Comme chaque opérateur dans l'interdomaine n'a qu'une vision locale de la topologie du réseau, il n'est donc pas capable de connaitre en temps réel ni le nombre de ses concurrents ou de ses acheteurs potentiels pour acheter ou revendre des routes, ni les caractéristiques  de ces routes. Pour prendre en considération l'ensemble de ces contraintes, nous modélisons ce problème par un jeu répété non cooperatif à information incomplète.
Le mécanisme de répétition que nous considérons ici, ne nous sert donc que pour la mise à jour des stratégies des joueurs, à la fin de chaque étape du jeu $t$. Une stratégie correspond à: comment un joueur fixe son intervalle de capacité à demander sur une route à destination donnée? 
Chaque joueur fixe son intervalle de capacité sur une route donnée dans le but de maximiser son bénéfice et aussi dans le souhait de gagner devant ses concurrents s'il est en concurrence avec d'autres joueurs pour l'achat de cette route.  
\\\indent 
Le réseau interdomaine est  représenté par un graphe non orienté $G(V, E)$ à $(n+1)$ n\oe{}uds, où $V$ et $E$ désignent respectivement les systèmes autonomes ASs (Autonomous System) et les liens du réseau. Pour chaque destination $d$,  nous considèrons un ensemble de joueurs $J=\{j_{1}, ... , j_{n}\}$ qui correspondent à l'ensemble des n\oe{}uds demandeurs de routes vers $d$ (la destination $d$ n'est donc pas considérée comme un joueur). Nous considérons que les actions de chaque joueur $j_{i}$ sont données par l'ensemble $A_{i}=\{a_{1}, ... , a_{m_{i}}\}$, où chaque composante de cet ensemble correspond à un intervalle de capacité $[cap\_min_{i,d},$ $cap\_max_{i,d}]$. Les valeurs $cap\_min_{i,d}$ et  $cap\_max_{i,d}$ représentent respectivement la capacité minimale et la capacité maximale que le joueur $j_i$ demande sur la route à destination $d$. Ces capacités sont discrètes, finies et bornées superieurement par la capacité totale $cap(i)$ du joueur $j_i$. Nous considérons qu'à chaque étape $t$ du jeu répété, l'ensemble des stratégies de chaque joueur $j_i$ est associé à un vecteur de probabilité $s_i^t$ sur l'ensemble de ses actions $A_i$. Soit $s_i^t=((s_{i,k}^t)_{k=1...m_i})$ tel que $s_{i,k}^t$ est la probabilité que le joueur $j_i$ joue l'action $k$ à l'étape $t$. Soit $p^t=(s_1^t, s_2^t, ..., s_n^t)$ un profil de stratégies d'un jeu constituant à l'étape $t$. Tout ce que connait chaque joueur se résume aux données suivantes:
\textbf{i)} S'il est traversé ou non par une certaine partie du trafic vers une destination donnée.
\textbf{ii)} Son bénéfice et les caractéristiques (prix, capacité, délai et disponibilité) des différents chemins qui lui sont annoncés par ses voisins.
Nous notons $Stra_i$ le processus local stratégique sur lequel se base le joueur $j_i$ pour calculer sa nouvelle distribution de probabilités à l'étape $t+1$ en fonction de celle à l'étape $t$. Au niveau de chaque jeu constituant, nous appliquons l'algorithme $\mathcal{L}$ qui consiste à suivre les étapes suivantes:
\begin{enumerate}
\item Chaque joueur $j_i$ fixe son intervalle de capacité selon  son vecteur de probabilités $s_i$.\vspace{-0.12cm}
\item  Chaque joueur $j_i$  choisit la route offerte de capacité maximum dans l'intervalle $[cap\_min_{i,d},$ $cap\_max_{i,d}]$ à prix minimum.\vspace{-0.12cm}
\item Chaque joueur $j_i$ déduit son bénéfice.\vspace{-0.12cm}
\item Chaque joueur $j_i$ met à jour son vecteur de probabilités selon son processus local $Stra_i$.\end{enumerate}
L'algorithme $\mathcal{L}$ est implementé localement au niveau de chaque joueur et calcule à chaque nouvelle étape (t+1) un nouveau profil de stratégies, donc $p^{t+1}=\mathcal{L}(p^t)$. 
Nous nous sommes basés sur une technique d'apprentissage similaire à celle utilisée dans \cite{MT06,Sastry94}. Cette technique appelée  
(
Linear Reward Inaction Algorithm
) est utilisée pour définir le processus stratégique $Stra_i$. 
Elle est basée sur la règle de mise à jour présentée par l'équation (\ref{eq1}):\vspace{-0.1cm}
\begin{equation}\label{eq1}
Stra_i(s_{i,k}^t) =s_{i,k}^{t+1}=
\left\{
  \begin{array}{ll}
    s_{i,k}^t-b*u_i^t*s_{i,k}^t~~si~~k\neq [cap\_min_{i,d}^t, cap\_max_{i,d}^t]\hbox{} \\\hbox{} \vspace{-0.3cm}\\
    s_{i,k}^t+b*u_i^t* \sum_{l\neq [cap\_min_{i,d}^t, cap\_max_{i,d}^t]} s_{i,l}^t~~sinon\hbox{} 
     \end{array}
\right.
\end{equation}\vspace{-0.2cm}
\begin{description}
\item[ ] $u_i^t=\frac{Benef_{i,d}^t-Benef\_min_{i,d}^t}{Benef\_max_{i,d}^t-Benef\_min_{i,d}^t}$: est l'utilité normalisée. Les variables $Benef\_max_{i,d}^t$ et $Benef\_min_{i,d}^t$ correspondent respectivement au bénéfice maximal et au bénéfice minimal du joueur $j_i$ depuis le début du jeu jusqu'à l'itération $t$ et $Benef_{i,d}^t$ est son bénéfice à l'itération $t$ sur la route vers $d$. $cap\_min_{i,d}^t$ et $cap\_max_{i,d}^t$ correspondent respectivement à la capacité minimale et à la capacité maximale demandées par $j_i$ à l'étape $t$ sur la route vers $d$. Le paramètre $b\in [0,~1]$ est un paramètre d'apprentissage qui permet de moduler la vitesse d'apprentissage des différents joueurs. Nous considérons que $\forall s_{i,k}^0\in s_i^0,~s_{i,k}^0\neq 0$, cette condition permet de donner à chaque action une chance d'être choisie au départ.
\end{description}\vspace{-0.5cm}
\section{Conclusion}\vspace{-0.3cm}
Un simulateur a été développé en C pour tester le modèle de jeu proposé. Les résultats de simulations obtenus montrent que dans le cas de topologies de réseaux simples avec une seule destination et en introduisant des stratégies simples, le modèle converge vers des états stables où un nombre important d'opérateurs sont satisfaits (taux de satisfaction intéressant) et le choix de chaque joueur converge vers une stratégie qui est un équilibre de Nash. 
\vspace{-0.2cm}
\bibliographystyle{plain}

\end{document}